\documentclass[a4paper,twocolumn]{article}

\usepackage[utf8]{inputenc}
\usepackage{ucs}
\usepackage[english]{babel}
\usepackage{amsmath,amsfonts,amssymb}
\usepackage{bm,braket,slashed}
\usepackage{feynmp-auto}
\usepackage[numbers,comma,square,compress]{natbib}
\usepackage[colorlinks=true,allcolors=blue]{hyperref}

\usepackage[pdftex,a4paper,hmargin=15mm,vmargin=30mm]{geometry}

\newcommand{\myemail}[1]{e-mail: \href{mailto:#1}{#1}}

\DeclareMathOperator{\Real}{Re}
\DeclareMathOperator{\Imag}{Im}

\begin{document}
\title{\textbf{Dispersion of Light Traveling through the Interstellar Space, Induced and Intrinsic Lorentz Invariance Violation}}

\author{\\\vspace*{-.6em}
\textbf{I.~Brevik}$\,^{a,}$\footnote{\myemail{iver.h.brevik@ntnu.no}}~,
\textbf{M.~Chaichian}$\,^{b,}$\footnote{\myemail{masud.chaichian@helsinki.fi}}
~and
\textbf{M.~Oksanen}$\,^{b,}$\footnote{\myemail{markku.oksanen@helsinki.fi}}\\
\vspace*{.8em}\\
$^a$\textit{Department of Energy and Process Engineering, Norwegian
University of Science and Technology,}\\\textit{N-7491 Trondheim, Norway}\\
$^b$\textit{Department of Physics, University of Helsinki, P.O.Box 64, FI-00014 Helsinki, Finland}
\vspace*{.6em}
}

\date{}
\maketitle

\begin{abstract}
Theoretical models and experimental observations suggest that gamma-ray
bursts (GRB) and high-energy neutrino bursts travelling through the
interstellar space may reach the Earth at different speeds.
We propose and study in details the mechanism i), which always exists, where GRB are slowed down due to the dispersion of light in the interstellar medium.
In addition to the standard media such as electrons and photons as CMB, we
consider the medium with invisible axions.
The amount of GRB delays in different media are calculated in details utilizing a novel technique in QFT by using the hitherto known or estimated densities of particles in the space without introducing any arbitrary parameter.
Previously, the GRB delays have been interpreted as a sign of Lorentz invariance
violation by modifying the dispersion relation of Special Relativity, which relates
the energy, the momentum and the mass of a particle, based on different
mechanisms ii), such as a stringy spacetime foam, coming from a
quantum gravity effect and using an adjustable parameter.
Obviously, all the above-mentioned mechanisms i) and ii) are induced (seeming) Lorentz invariance violations but not an intrinsic (genuine) one.
The amount of GRB delay due to the two aforementioned interpretations can be distinguished by observing the time of arrival of light with different frequencies. Namely, dispersion of light i) predicts that the higher energy GRB arrive the Earth earlier, while in the other interpretations ii), they arrive later. We notice that the needed amount for delay due to the dispersion of light shall have the potential power to shed additional light on the microstructure of interstellar media with respect to the densities of constituent particles and the origins of their sources.
Finally, we indicate the ways to detect the \emph{intrinsic} Lorentz invariance violation and to interpret them theoretically.
\end{abstract}

\section{Introduction}
The propagation of gamma-ray bursts (GRB) in the interstellar medium
is a topic that has attracted considerable interest.
GRB are highly energetic and diverse events, which
are thought to be produced by violent stellar processes, in particular
the supernovas and mergers of binary neutron stars. Those events may also
produce high-energy cosmic rays and consequently bursts of high-energy
neutrinos \cite{Waxman:1997ti,Vietri:1998jr}. However, coincident GRB
and neutrino bursts have not been observed. The possibility that the
neutrino burst could be shifted in time with respect to the GRB is
under active study.
The time window $\tau=t_{GRB}-t_{\nu}$ between the arrival times of a
GRB, $t_{GRB}$, and a neutrino burst, $t_{\nu}$, could vary between seconds
or several days. References about this can be found in the report
\cite{Adrian-Martinez:2016xij} in connection with the ANTARES neutrino
telescope. Experimental data taken between 2007 and 2012 were analyzed;
gamma-ray energies analyzed were up to 100 TeV.
Assuming that a GRB and the corresponding neutrino burst
are produced at the same time, a significant delay $\tau$ would indicate
that the electromagnetic and neutrino signals have travelled at different speeds.
Note, however, that the recent experimental studies show only faint neutrino
signals associated with GRB \cite{Adrian-Martinez:2016xij,Aartsen:2020eof},
and hence the observed delays may be inaccurate.

Theoretical interpretations of the GRB delay phenomena have proved to
be challenging. Violation of Lorentz invariance at very high energies
in the form of modifying the dispersion relation has been
considered as a potential interpretation of the delay of high-energy GRB
\cite{AmelinoCamelia:1997gz,Ellis:1999sd,Jacob:2006gn,Jacob:2008bw}.
This approach is motivated by various approaches to quantum gravity,
as quantum-gravitational fluctuations may lead to a nontrivial
refractive index \cite{Ellis:2008gg}. The stringy spacetime foam is a
realization of such an effect \cite{AmelinoCamelia:1996pj}.
Those are examples of induced violation of Lorentz invariance.

We consider a mechanism within the standard physics. Namely, in the
presence of media which interact with photons, the dispersion of light
always occurs what induces changes in the speed of light from its
value $c$ in the vacuum.
Along this line we study the dispersion of light in several interstellar media
and assess the produced GRB delay when photons and neutrinos are assumed to
be emitted from the same source and simultaneously.
Neither an electron-positron plasma nor a photon medium can account for
significant GRB delays, taking the observed or the assumed densities of
electrons and photons are too low to slowdown the high-energy photons sufficiently. By a plasma here it is meant a medium where the wave-length
of the incoming light is much smaller than the free path of the
constituent particles of the medium.
Therefore, we study the properties of an axionic medium. Axions are
pseudoscalar particles that may both provide a solution to the strong
CP problem and as well be a candidate for cold dark matter (CDM).
They are not electrically charged but can still interact with photons.
Axion electrodynamics has been studied actively and it is connected
to topological insulators
\cite{Wilczek:1987mv,Basar:2012gm,Martin-Ruiz:2015skg,Fukushima:2019sjn}.
While the original axion model was ruled out by observations, a new version
of the axion, which is called the \emph{invisible axion}, is consistent
with laboratory experiments and astrophysical observations \cite{Sikivie:2020zpn}.
Therefore, an axion medium is a plausible cosmic medium that would have
an effect on the propagation of light from distant galaxies. We derive
the dispersion relation in an axion medium and assess its effect on the
GRB delay.

Our approach is based on the quantum field theory and the optical
theorem. We derive the value of of the plasma frequency $\omega_p$,
i.e., the term contributing to the change in the speed of light.
To our knowledge such a derivation and the result has not been given before.
Then the corresponding dispersion relation in this medium and the
produced GRB delay are derived.Our estimates show that the interaction
between GRB and axions is too weak for producing a significant delay
between the gamma-ray and neutrino signals.

To make our physical picture clear: we assume that the photons and
neutrinos are emitted from the same source at the same time, and
calculate the value of $\tau$ in all plausible media in the
interstellar space and for a few selected values of the incoming GRB
frequency. The dispersive effect indicates that the refractive index
will be less than unity, corresponding to the phase velocity being
superluminal, while the group velocity is subluminal.
We neglect the dissipation effect as this is expected to be weak.
At the highest photon frequencies where the permittivity is
very close to unity, the photon group velocity is slightly lower than
$c$, so that for these frequencies the delay necessarily has to
be the least.

Notice that the interaction of neutrinos with any interstellar medium
is extremely weak and hence the dispersion of neutrinos is negligible.
Secondly, while the neutrinos are massive and oscillating, the effect on
the speed of high-energy neutrinos is very small. Consider a GRB with
photon energy 1 TeV and neutrinos with the same energy,
$E=1\;\mathrm{TeV}$. The dispersion relation $E^2=p^2c^2+m^2 c^4$ gives
the speed of the neutrinos as $v_\nu=\frac{dE}{dp}\approx c(1-d_\nu)$,
where $d_\nu=\frac{m^2c^4}{2E^2}$. Averaging over 3 neutrinos, $\langle
m^2c^4\rangle = (1/3) (0.1\,\mathrm{eV})^2$, where the masses are
estimated with the heaviest neutrino mass. The speed of neutrinos is
given by $d_\nu=1.7 \times 10^{-27}$. Thus the delay compared to a
signal travelling at the speed $c$ would be measured in nanoseconds even
for signals from furthest galaxies: $\tau=D\times d_\nu/c
\lesssim10^{-9}\;\mathrm{s}$, using a maximal travelling distance
$D=1.4\times 10^{26}\;\mathrm{m}$ (effective distance to farthest
galaxies around $z=10$). That is a negligible time delay compared to
the GRB delays searched in experiments. Some theories of neutrino
production in GRB actually predict neutrinos with even higher energy of
order $10^{2}$--$10^{7}$ TeV \cite{Waxman:1997ti}, which means $v_{\nu}$
might be even closer to $c$. These arguments justify to take $v_{\nu}=c$
in our estimates.

The time delay generated by dispersion of light in a medium in its form
is essentially different from those caused by the induced Lorentz
invariance violations due to quantum gravity effects and in particular a
spacetime foam: in the latter cases, the speed of light becomes less for higher
energy photon and thus the time delays compared with neutrino bursts
grows, a distinguishable effect opposite to the effect of light
dispersion in all the media. We shall not consider those (induced)
Lorentz invariance breakings in detail here, although some remarks
will be given in Sec.~\ref{secLIV}. We mention though that by assuming
a spacetime foam as the cosmic medium, it has been found
\cite{Ellis:2008gg,Mavromatos:2010pk} that the delay
time is longest for the more energetic photons. This points to an
important observation: if one considers one single GRB, the dispersive
theory predicts the highest frequencies to move faster, while in the
Lorentz violation theory these frequencies will move slower. An
experimental test of these theories is thus possible.

Finally, we will mention some remarks and remind of some
\emph{intrinsic} Lorentz invariance violation effects.

The Minkowski metric is defined with the signature
$(\eta_{\mu\nu})=\mathrm{diag}(+1,-1,-1,-1)$.

\section{GBR delay in usual cosmic media}
\subsection{Dispersion relation and plasma frequency}
A plasma can support both longitudinal and transverse waves.
We are interested in transverse waves.
Dispersion relation for light in a plasma is \cite{Langmuir}
\begin{equation}
\omega^2=c^2\bm{k}^2+\omega_p^2, \label{dispersionrelation}
\end{equation}
where $\omega_p$ is the plasma frequency, which is due to the plasma
oscillations called Langmuir waves, derived from the classical Maxwell equations.
The angular frequency is also given as
$\omega=\bm{v}(\hat{\bm{k}})\cdot\bm{k}=\frac{c|\bm{k}|}{n}$,
where $n$ is the refraction index and $\bm{v}=\frac{c}{n}\hat{\bm{k}}$
is the phase velocity, where
$\hat{\bm{k}}=\frac{\bm{k}}{|\bm{k}|}$. Thus the refraction index is
related to the plasma frequency as
\begin{equation}
n^2=1-\frac{\omega_p^2}{\omega^2}. \label{refractionindex}
\end{equation}
In an isotropic medium, $\omega=\omega(|\bm{k}|)$ and
$\omega_p=\omega_p(|\bm{k}|)$, group velocity is parallel to phase
velocity, $\bm{v}_g=v_g\hat{\bm{k}}$, with the magnitude given as
\begin{equation}
v_g=\frac{\partial\omega(|\bm{k}|)}{\partial|\bm{k}|}
=c\frac{c|\bm{k}|+\frac{1}{2c}
\frac{\partial\omega_p^2}{\partial|\bm{k}|}}
{\sqrt{\omega_p^2+c^2\bm{k}^2}}.
\end{equation}
Since superluminal propagation of energy and information is forbidden,
we demand that the plasma frequency $\omega_p(|\bm{k}|)$ must satisfy
the following condition for any $\bm{k}$,
\begin{equation}
\frac{c|\bm{k}|+\frac{1}{2c}
\frac{\partial\omega_p^2}{\partial|\bm{k}|}}
{\sqrt{\omega_p^2+c^2\bm{k}^2}}\le1. \label{pf-condition}
\end{equation}
We should note that a superluminal group velocity does not imply
violation of special relativity, since in such a case the signal
velocity is not equal to group velocity \cite{Milonni}. We still consider only cases with $v_g\le1$ so that the group velocity can be taken as the signal velocity.
If the plasma frequency is independent of $\bm{k}$, which is
the case in a usual plasma with charged particles, and the photon
momentum is large compared to the plasma frequency,
$c^2\bm{k}^2\gg\omega_p^2$, we obtain that group
velocity is only slightly lower than $c$,
\begin{equation}
v_g=\frac{\partial\omega(|\bm{k}|)}{\partial|\bm{k}|}
\cong c\left( 1-\frac{\omega_p^2}{2\omega^2} \right)\equiv c(1-d),
\label{groupvelocity}
\end{equation}
where $d\equiv\frac{\omega_p^2}{2\omega^2}\ll1$.

\subsection{Electron-positron plasma}
For an electron-positron plasma, the plasma frequency is obtained from classical
electrodynamics \cite{Langmuir,Jackson,CMRT-ED} as (in SI units)
\begin{equation}
\omega_p^2=\frac{Ne^2}{\epsilon_0m_e}, \label{pf.electron}
\end{equation}
where $N$ is the number density of electrons, $e^2$ is the square of
the electric charge of the electron, $\epsilon_0$ is the permittivity
of vacuum, and the mass of the electron is $m_e = 9.11\times
10^{-31}\;\mathrm{kg}=0.511\;\mathrm{MeV}/c^2$.
The same result \eqref{pf.electron} is obtained from quantum field
theory, which is shown in Appendix~\ref{appendix}.
The lightest charged particles have the greatest effect on the
dispersion of light. The number density of electrons $N$ is left
unspecified for now.

From \eqref{pf.electron} we get the plasma frequency
$\omega_p= 56\sqrt{\bar{N}}$ rad/s, where $\bar{N}$ denotes
the number density of electrons as a dimensionless quantity measured in
SI units: $\bar{N}=\frac{N}{[N]}=N\times\mathrm{m}^3$, where
$[N]=\mathrm{m}^{-3}$.
The angular frequency for the gamma-ray energy
$E$ is $\omega= 1.5\times\bar{E}\times 10^{27}$ rad/s, where
$\bar{E}=\frac{E}{\mathrm{GeV}}$ is gamma-ray energy in GeV units.
We obtain the group velocity \eqref{groupvelocity} of the gamma-ray as
\begin{equation}\label{v_GRB.electron}
v_{GRB}=c\left(1-0.7\times 10^{-45}\times\bar{N}/\bar{E}^{2}\right).
\end{equation}

Assume now that $D(z)$ is the effective distance travelled by the
photons taking into account the expansion of the Universe. It is
defined as \cite{Adrian-Martinez:2016xij}
\begin{equation}
D(z)=\frac{c}{H_0}\int_0^z
\frac{(1+z')dz'}{\sqrt{\Omega_m(1+z')^3+\Omega_\Lambda}},
\end{equation}
where $z$ is the redshift, $H_0$ the present Hubble parameter, and
$\Omega_m$ and $\Omega_\Lambda$ are the standard symbols for the
relative matter and dark energy densities. (For an alternative to this
method, see \cite{Dodin:2010zz}.) In principle, the photon
transit time is
\begin{equation}
t=\frac{D(z)}{v_{GRB}}. \label{tid}
\end{equation}
In order to evaluate this, one ought to include the $z$-dependence of
$v_{GRB}$ due to the varying density of the plasma. However,
we shall ignore the $z$-dependence of $N$ and $v_{GRB}$ for now, since
we are only dealing with some estimates. The time delay between two
signals travelling at the speeds \eqref{groupvelocity} and $c$ is
obtained as
\begin{equation}
\tau=\frac{D\times d}{c(1-d)}\cong\frac{D\times d}{c}. \label{timedelay}
\end{equation}

For the electrons \eqref{v_GRB.electron} we obtain the GRB delay
\begin{equation}
\tau=0.7\times 10^{-45}\times\frac{\bar{N}}{\bar{E}^{2}}\times
\frac{D}{c}.
\end{equation}
As a first estimate, we may consider galaxy filaments, which are the
greatest structures in the universe. The size of the largest filaments
is measured in gigaparsecs. While we do not know the electron
density in the largest known filaments, we can estimate it with the
known electron density of closer filaments. Hence we use the electron
density of galaxy filaments around $z=0.1$ \cite{FraserMcKelvie:2011ms}:
$N_e=(4.7\pm0.2)\times10^{-4}h_{100}^{1/2}\;\mathrm{cm}^{-3}
\approx4\times10^2\;\mathrm{m}^{-3}$. The effective distance is
chosen as $D=3\;\mathrm{Gpc}=9\times10^{25}\;\mathrm{m}$.
The delay produced by electrons for a GRB propagating through such a
structure is estimated as
\begin{equation}
\tau=\frac{0.8\times10^{-25}\;\mathrm{s}}{\bar{E}^{2}},\label{delay.ep}
\end{equation}
which is negligible for high-energy photons, in particular for GRB
photons with $\bar{E}\ge1$.

We conclude that the dispersive properties of an electron gas are not
significant enough to account for measurable GRB time delays. In fact,
even the delay of neutrinos caused by the masses of the neutrinos is
much longer than the delay due to dispersion of light in the
cosmic electron medium.

\subsection{Photon medium}
Dispersion of light in photon medium also produces a GRB delay that is
far too small to explain the observed delay. Light-on-light scattering
has been studied in particle accelerator experiments in great detail.
While direct observation of light-on-light scattering is difficult to
achieve in particle accelerators \cite{dEnterria:2013zqi}, evidence for
it is increasing and it also used for the search of axion particles
\cite{Sirunyan:2018fhl,Aad:2020cje}.
When dealing with strong fields this process has attracted considerable
interest and new types of experiments have been proposed; cf., for
instance, Refs.~\cite{King:2013am,Marklund:2010}.

We obtain from light-on-light scattering
\cite{Euler:1936,Heisenber:1936,Akhiezer:1937,Karplus:1951,Berestetsky:1982} (see Appendix~\ref{appendix.photon})
\begin{equation}
\omega_p^2=\mathrm{const.}\times\frac{N_{\gamma}e^4}{\omega},
\end{equation}
where $N_{\gamma}$ is the number density of photons
and $\omega=\sqrt{s}/2$ in terms of the invariant $s$.
According to the Planck data on the CMB (Cosmic Microwave Background)
radiation, which constitute the majority of photons in the Universe, we
have the number density of photons
$N_{\gamma}=(4$--$5)\times10^{8}\;\mathrm{m}^{-3}$. The GRB delay
produced by CMB is estimated as
\begin{equation}
\tau_{\mathrm{CMB}}=\frac{4\times10^{-41}\;\mathrm{GeV}^3}{s^{3/2}}\times
\frac{D}{c}=\frac{4\times10^{-23}}{\bar{E}^{3/2}}\times\frac{D}{c}. \label{d}
\end{equation}
For GRB originating from the farthest galaxies the delay produced by
the interaction with CMB is
$\tau_{\mathrm{CMB}}=2\times10^{-5}\;\mathrm{s}/\bar{E}^{3/2}$;
e.g. for a gamma-ray energy $E=100\;\mathrm{GeV}$ the delay is
$\tau_{\mathrm{CMB}}=2\times10^{-8}\;\mathrm{s}$. This is a very short
delay but still much longer than the delay produced by electrons \eqref{delay.ep}.

The extra-galactic background light (EBL) is the second, after the CMB, most abundant part of the photon medium in the Universe.
With the data as given in \cite{Ajello:2018sxm,Aliu:2008ay} (see also
\cite{HESS:2005ohe}), we obtain
the photon number density $N_{\mathrm{EBL}}=10^{4}\;\mathrm{m}^{-3}$, and
with a typical EBL photon energy of $1\;\mathrm{eV}$, we obtain the delay
$\tau_{\mathrm{EBL}}=10^{-32}/\bar{E}^{-3/2}\times(D/c)$.
Thus, the number of CMB photons is several orders of magnitude larger than the number of EBL photons and the same is with
$\tau_{\mathrm{CMB}}$ compared with $\tau_{\mathrm{EBL}}$.

For high-energy light propagating in the Universe, there also appears the electromagnetic cascades due to the electron-positron pair production,
what adds an additional contribution to the ordinary electromagnetic background.
However, this process goes through a higher order in the electromagnetic
coupling constant and its contribution to the delay $\tau$ can be neglected.

Thus the dispersion on background photons does not produce a
significant GRB delay.

\section{Axions and their effect on the propagation of gamma-rays}
\label{sec:axion}
Since the GRB delay produced by a charged plasma was found to be
proportional to the inverse of the particle mass, $\tau\propto m^{-1}$,
it becomes natural to look for particles of much lower mass than the
electron. As mentioned above, we will consider a model where the
dark matter consists of axions.
However, since an electrically charged axion is not consistent with
experiments and observations, the plasma frequency formula for charged
plasma \eqref{pf.electron} is no longer valid. Nevertheless, with the
vivid activity in axion electrodynamics
\cite{Wilczek:1987mv,Basar:2012gm,Martin-Ruiz:2015skg,Fukushima:2019sjn}
with its connection to topological insulators (for experiment, see
CERN Axion Solar Telescope), the assumption of an axionic plasma with
its coupling to photons seems quite appropriate.
The characteristic axion mass for the QCD axion experiments is about
$m_{a}=10^{-5} \rm{ eV}/c^2$ \cite{Sikivie:2020zpn}. However, the
axion mass may be much smaller: a satisfactory agreement with
constraints has been reported when the axion mass lies in the interval
$10^{-18}\;\mathrm{eV}<m_a<10^{-28}\;\mathrm{eV}$
\cite{Gorghetto:2021fsn}.

The effective coupling between the axion and two photons is given by the
interaction Lagrangian \cite{Sikivie:2020zpn,Halverson:2019cmy}
\begin{equation}
\mathcal{L}_{a\gamma\gamma}=-\frac{1}{4}g_\gamma
\frac{\alpha}{\pi}\frac{1}{f_a}a(x)F^{\mu\nu}(x)\tilde{F}_{\mu\nu}(x),
\label{L_int}
\end{equation}
where $\alpha$ is the fine structure constant and
\begin{equation}
g_\gamma = \frac{1}{2}\left( \frac{\mathcal{N}_e}{\mathcal{N}}
-\frac{5}{3}-\frac{m_d-m_u}{m_d+m_u} \right).
\end{equation}
Here $\mathcal{N}$ and $\mathcal{N}_e$ are respectively the color
anomaly and electromagnetic anomaly, $m_d$ and $m_u$ are the quark
masses, $f_a$ is the axion decay constant, and $a(x)$ is the axion
field. The electromagnetic field strength tensor is
$F_{\mu\nu}=\partial_\mu A_\nu-\partial_\nu A_\mu$, and its dual
$\tilde{F}^{\mu\nu}=\frac{1}{2}\epsilon^{\mu\nu\rho\sigma}
F_{\rho\sigma}$. We define the effective coupling constant $g$ as
\begin{equation}
g = g_\gamma \frac{\alpha}{\pi}\frac{1}{f_a}.
\end{equation}
If the interaction Lagrangian is written in terms of the electric
$\bm{E}$ and the magnetic field $\bm{B}$, we would have
$\mathcal{L}_{a\gamma\gamma}=-ga(x)\bm{E}\cdot\bm{B}$.
Here we adopt a system of units with Heaviside--Lorentz electromagnetic
units and $\hbar=c=1$.

First we note that the optical properties of axion backgrounds have
been explored in \cite{McDonald:2019wou,McDonald:2020why} and
references therein by means of classical axion electrodynamics.
The issues concerning GRB have not been considered.
The group velocity of light in the presence of the axion field and
without charged plasma is given as
$v_g=1+g^2(\partial_\mu a)^2/8k_0^2$ \cite{McDonald:2019wou}.
If the vector $\partial_\mu a$ is timelike, $(\partial_\mu a)^2>1$, the
group velocity is not the velocity at which information propagates, and
in such cases one should instead use true signal velocity
\cite{Milonni}. In order to apply the aforementioned formula for $v_g$
to the problem of GRB delay one would need to obtain the value of
$(\partial_\mu a)^2$ along the route of GRB, or relate $(\partial_\mu
a)^2$ to the density of axions. We shall not do that here. Instead we
use an approach that is based on quantum field theory.

We would like to mention that while the aim of \cite{McDonald:2019wou} has been to follow the trajectory of the electromagnetic  field to obtain the polarisation vector and the possibility of birefrigence effect, which can occur only in a chiral medium, e.g. in an axionic one, our aim has been to find out the group velocity of GRB travelling through different media such as electron-positron, axion and CMB, each separately, and the delay time for each of them. For that purpose we have considered the dispersion of unpolarized light, and the emerging time delays depending on the photon energy, what is the relevant case for the present GRB experiments. The unpolarized case is obtained in the quantum field theoretical derivation of the scattering amplitude by averaging over the initial polarization states and summing over the final polarization states. In this way the effect of the gyration vector cancels out and no birefringence appears. In future, when the experimental facilities and detectors will have the precision of measuring the polarised GRB bursts with different delay times, the analysis performed in \cite{McDonald:2019wou} will be most useful.

We derive the dispersion equation \eqref{dispersionrelation} for
unpolarised coming and detected lights in the axionic matter.
A quantum field theoretical calculation of the plasma frequency
$\omega_p^2$ is given next. The calculation holds for high-energy photons,
when the photon energy is much higher than the mass of the axion $m_{a}$.
A calculation like this has to our knowledge not been given before,
neither in classical electrodynamics nor in quantum field theory.

\subsection{Derivation of the dispersion relation of light in axion
medium}
\subsubsection{Calculation of the scattering amplitude}
We use the technique illustrated in Appendix~\ref{appendix}. The
effective interaction Lagrangian \eqref{L_int} for invisible axion is
\begin{equation}
\mathcal{L}_{a\gamma\gamma}=-\frac{1}{4}ga F_{\mu\nu}\tilde{F}^{\mu\nu}
=-\frac{1}{2}ga\epsilon^{\mu\nu\rho\sigma}\partial_\mu A_\nu
\partial_\rho A_\sigma.
\end{equation}
The Feyman rule for the interaction vertex is read from the interaction
Lagrangian as
$\frac{i}{2}g\epsilon^{\mu\nu\rho\sigma}k^{(1)}_{\mu}k^{(2)}_{\rho}$,
where $k^{(1)}$ and $k^{(2)}$ are the four-momenta of the two photons
and the free indices $\nu$ and $\sigma$ correspond to the first and
second photons, respectively.

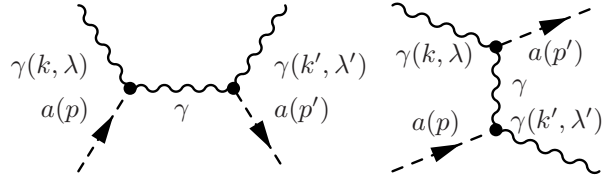
\begin{figure}[ht]
\begin{center}
\begin{fmffile}{axion-diagrams}
\begin{fmfgraph*}(34,22)
\fmfleft{i1,i2} \fmfright{o1,o2}
\fmf{scalar,label=$a(p)$,label.side=left}{i1,v1}
\fmf{photon,label=$\gamma(k,,\lambda)$,label.side=right}{i2,v1}
\fmf{photon,label=$\gamma$}{v1,v2}
\fmf{scalar,label=$a(p')$,label.side=left}{v2,o1}
\fmf{photon,label=$\gamma(k',,\lambda')$,label.side=right}{v2,o2}
\fmfdot{v1,v2}
\end{fmfgraph*}
\hspace{5mm}
\begin{fmfgraph*}(34,22)
\fmfleft{i1,i2} \fmfright{o1,o2}
\fmf{scalar,label=$a(p)$,label.side=left}{i1,v2}
\fmf{photon,label=$\gamma(k,,\lambda)$,label.side=right,label.dist=7}{i2,v1}
\fmf{photon,label=$\gamma$}{v1,v2}
\fmf{photon,label=$\gamma(k',,\lambda')$,label.side=left,label.dist=7}{v2,o1}
\fmf{scalar,label=$a(p')$,label.side=right}{v1,o2}
\fmfdot{v1,v2}
\end{fmfgraph*}
\end{fmffile}
\end{center}
\caption{Feynman diagrams for scattering of photon on axion.}
\label{fig.a}
\end{figure}

We consider scattering of photon on axion,
$\gamma+a\rightarrow\gamma+a$, at tree level.
The scattering amplitude $\mathcal{M}$ is a sum of the two diagrams
in Fig.~\ref{fig.a}.
\unitlength = 1mm 

First we consider the $s$-channel contribution in detail (first diagram
in Fig.~\ref{fig.a}). Ingoing photon and axion have four-momenta $k$ and
$p$, respectively. The resulting virtual photon propagates with momentum
$k+p$. Outgoing momenta are primed $k'$ (photon) and $p'$ (axion), and
four-momentum is conserved $k+p=k'+p'$. The contribution of the
$s$-channel diagram to the scattering amplitude in Feynman gauge is
\begin{equation}
\begin{split}
i\mathcal{M}_{(s)}&=\frac{i}{2}g\epsilon^{\alpha\beta\gamma\delta}
(k_\alpha+p_\alpha) \epsilon^{*}_\beta(\bm{k}',\lambda')k'_\gamma
\frac{-ig_{\delta\sigma}}{(k+p)^2}\\
&\times\frac{i}{2}g\epsilon^{\mu\nu\rho\sigma}k_{\mu}
\epsilon_{\nu}(\bm{k},\lambda) (k_{\rho}+p_{\rho}),
\end{split}
\end{equation}
where at the right-hand side of the propagator we have the first vertex.
That is simplified as
\begin{equation}
\mathcal{M}_{(s)}=-\frac{3!g^2}{4}
\frac{(k_\mu+p_\mu)\epsilon^{*}_{\nu}(\bm{k}',\lambda')k'_{\rho}
k^{[\mu}\epsilon^{\nu}(\bm{k},\lambda)p^{\rho]}}{(k+p)^2}
\end{equation}
using
\begin{equation}
\epsilon^{\mu\nu\rho\sigma}k_{\rho}k_{\sigma}=0
\end{equation}
and
\begin{equation}
\begin{split}
\epsilon^{\alpha\beta\gamma\delta}g_{\delta\sigma}
\epsilon^{\mu\nu\rho\sigma}
&=-\delta^{\alpha\beta\gamma}_{\mu'\nu'\rho'}
g^{\mu'\mu}g^{\nu'\nu}g^{\rho'\rho}\\
&=-3!\delta^{[\alpha}_{\mu'}\delta^{\beta}_{\nu'}\delta^{\gamma]}_{\rho'}
g^{\mu'\mu}g^{\nu'\nu}g^{\rho'\rho}.
\end{split}
\end{equation}
Then we write out the antisymmetrization of the indices $\mu\nu\rho$
and use the gauge condition $k_\mu\epsilon^{\mu}(\bm{k},\lambda)=0$,
\begin{equation}
\begin{split}
\mathcal{M}_{(s)}&=-\frac{3g^2}{2}\frac{1}{(k+p)^2}
\bigl[ (k_\mu+p_\mu)k^{\mu}\\
&\times\epsilon^{*}_{\nu}(\bm{k}',\lambda')\epsilon^{\nu}(\bm{k},\lambda)k'_{\rho}p^{\rho}\\
& -(k_\mu+p_\mu)k^{\mu}\epsilon^{*}_{\nu}(\bm{k}',\lambda')p^{\nu}
k'_{\rho}\epsilon^{\rho}(\bm{k},\lambda)\\
& -p_\mu\epsilon^{\mu}(\bm{k},\lambda)
\epsilon^{*}_{\nu}(\bm{k}',\lambda')k^{\nu}k'_{\rho}p^{\rho}\\
& +(k_\mu+p_\mu)p^{\mu}\epsilon^{*}_{\nu}(\bm{k}',\lambda')k^{\nu}
k'_{\rho}\epsilon^{\rho}(\bm{k},\lambda)\\
& +p_\mu\epsilon^{\mu}(\bm{k},\lambda)
\epsilon^{*}_{\nu}(\bm{k}',\lambda')p^{\nu}k'_{\rho}k^{\rho}\\
& -(k_\mu+p_\mu)p^{\mu}\epsilon^{*}_{\nu}(\bm{k}',\lambda')
\epsilon^{\nu}(\bm{k},\lambda)k'_{\rho}k^{\rho} \bigr].
\end{split}
\end{equation}
Now use $k^2=0$ for the initial photon and $p^2=m_a^2$ for the axion:
\begin{equation}
\begin{split}
\mathcal{M}_{(s)}&=-\frac{3g^2}{2}\frac{1}{(2k\cdot p+m_a^2)} \bigl[
\bigl( (k\cdot p)(k'\cdot p)\\
&-(k\cdot p+m_a^2)(k'\cdot k) \bigr)
\epsilon^{*}_{\mu}(\bm{k}',\lambda')\epsilon^{\mu}(\bm{k},\lambda)\\
&+\bigl( (k\cdot p+m_a^2)k^{\mu}k'_{\nu}
-(k'\cdot p)k^{\mu}p_{\nu} \\
&-(k\cdot p)p^{\mu}k'_{\nu} +(k'\cdot k)p^{\mu}p_{\nu} \bigr)\\
&\times\epsilon^{*}_{\mu}(\bm{k}',\lambda')\epsilon^{\nu}(\bm{k},\lambda)
\bigr]. \label{Ms}
\end{split}
\end{equation}

When the momenta of the initial and final photons are parallel (with an
angle $\theta=0$ between $\bm{k}$ and $\bm{k}'$), we have $k\cdot k'=0$,
and we also have $k'_{\mu}\epsilon^{\mu}(\bm{k},\lambda)=0$ and
$k^{\mu}\epsilon^{*}_{\mu}(\bm{k}',\lambda')=0$ from the gauge
condition. Thus only the first term of \eqref{Ms} survives in forward
scattering. Furthermore, since the momenta of the initial and final
photons are parallel, their polarization can be described with same
vectors, which are taken to be orthonormal,
$\epsilon^{*}_{\mu}(\bm{k},\lambda')\epsilon^{\mu}(\bm{k},\lambda)
=-\delta_{\lambda'\lambda}$. Therefore for $\theta=0$ the amplitude
\eqref{Ms} becomes
\begin{equation}
\mathcal{M}_{(s)}(0)=\frac{3g^2}{2}\delta_{\lambda'\lambda}
\frac{(k\cdot p)(k'\cdot p)}{(2k\cdot p+m_a^2)}.
\end{equation}
Then we assume that axions are very cold so that their linear
three-momenta $\bm{p}$ are very small, $m_a\gg p^i$ and $k_0\gg p^i$
($i=1,2,3$). In the scattering amplitude, we can approximate
$k\cdot p =k_0\sqrt{m_a^2+\bm{p}^2}-\bm{k}\cdot\bm{p}\cong
k_0m_a$. This approximation becomes exact in the rest frame of the
initial axion, $\bm{p}=0$, which we shall use as the ``laboratory
frame'' of our calculation.
Furthermore, we obtain from conservation of four-momentum that
$(k+p-k')^2=p^{\prime 2}$, where the left-hand side is
$2k\cdot p+m_a^2-2k'_{\mu}p^{\mu}$ in forward
scattering and the right-hand side is $m_a^2$. This implies $k'_0=k_0$.
We obtain
\begin{equation}\label{Ms0}
\mathcal{M}_{(s)}(0)=\frac{3g^2}{2}\delta_{\lambda'\lambda}
\frac{k_0^2 m_a}{(2k_0+m_a)}.
\end{equation}

The $u$-channel contribution (second diagram in Fig.~\ref{fig.a}) is
similar
to the $s$-channel but with the four-momentum of the virtual photon
$k+p$ replaced by $k-p'$,
\begin{equation}
\begin{split}
i\mathcal{M}_{(u)}&=\frac{i}{2}g\epsilon^{\alpha\beta\gamma\delta}
(k_\alpha-p'_\alpha) \epsilon^{*}_\beta(\bm{k}',\lambda')k'_\gamma
\frac{-ig_{\delta\sigma}}{(k-p')^2}\\
&\times\frac{i}{2}g\epsilon^{\mu\nu\rho\sigma}k_{\mu}
\epsilon_{\nu}(\bm{k},\lambda) (k_{\rho}-p'_{\rho}).
\end{split}
\end{equation}
Thus the $u$-channel contribution to the forward scattering amplitude
is obtained as
\begin{equation}
\mathcal{M}_{(u)}(0)=\frac{3g^2}{2}\delta_{\lambda'\lambda}
\frac{(k\cdot p')(k'\cdot p')}{(-2k\cdot p'+m_a^2)}.
\end{equation}
Using the conservation of four-momentum, we get
\begin{equation}\label{Mu0}
\mathcal{M}_{(u)}(0)=\frac{3g^2}{2}\delta_{\lambda'\lambda}
\frac{k_0^2 m_a}{(-2k_0+m_a)}.
\end{equation}

The scattering amplitude for $\theta=0$ is
\begin{equation}
\begin{split}
\mathcal{M}(0)&=\mathcal{M}_{(s)}(0)+\mathcal{M}_{(t)}(0)\\
&=\frac{3g^2}{2}\delta_{\lambda'\lambda}\left( \frac{k_0^2
m_a}{(2k_0+m_a)}
+\frac{k_0^2 m_a}{(-2k_0+m_a)} \right),
\end{split}
\end{equation}
and its square for unpolarized photons is
\begin{multline}
\frac{1}{2}\sum_{\lambda,\lambda'}|\mathcal{M}(0)|^2
=\left(\frac{3g^2}{2}\right)^2\left| \frac{k_0^2 m_a}{(2k_0+m_a)}
+\frac{k_0^2 m_a}{(-2k_0+m_a)} \right|^2\\
=\left(3g^2\right)^2 \left|\frac{k_0^2m_a^2}{-4k_0^2+m_a^2}
\right|^2.
\end{multline}
The amplitude diverges at $k_0=\frac{1}{2}m_a$, which would generally
require regulation. However, here we are concerned with the high-energy
case $k_0\gg m_a$.

The differential cross section for unpolarized photons is
\begin{equation}
d\sigma=\frac{1}{2k_0}\frac{1}{2p_0}
\left( \frac{1}{2}\sum_{\lambda,\lambda'}|\mathcal{M}|^2 \right)
d\mathrm{Lips},
\end{equation}
where the relative velocity of the initial particles is $c=1$ (in any
frame) and the Lorentz invariant two-body phase space is the same one
as in the case of photon-electron scattering
\eqref{dLips}--\eqref{dLips.2}.
The differential cross section is written as
\begin{equation}
d\sigma=\frac{k'_0}{64\pi^2k_0p_0p'_0}
\left( \frac{1}{2}\sum_{\lambda,\lambda'}|\mathcal{M}|^2 \right)
d\Omega.
\end{equation}

We now choose the rest frame of the initial axion and specialize to
forward scattering $\theta=0$. We have seen that $k'_0=k_0$ and
$p_0=p'_0=m_a$, and hence the differential cross section is
\begin{equation}
\begin{split}
d\sigma(0)&=\frac{1}{64\pi^2m_a^2}
\left( \frac{1}{2}\sum_{\lambda,\lambda'}|\mathcal{M}(0)|^2 \right)
d\Omega\\
&=\left( \frac{3g^2}{8\pi} \right)^2 \left|
\frac{k_0^2m_a}{-4k_0^2+m_a^2} \right|^2 d\Omega.
\end{split}
\end{equation}
According to \eqref{fsa.abs} the absolute value of the forward
scattering amplitude is given as
\begin{equation}\label{fsa.axion}
|f(0)|=\frac{3g^2}{8\pi}\frac{\omega^2m_a}{|4\omega^2-m_a^2|},
\end{equation}
where the energy of the photon is given by its angular frequency,
$k_0=\omega$. Unlike in the case of electron-positron plasma, the forward
scattering amplitude depends on the photon energy.

Finally, we note that in the dispersion relation
\eqref{dispersionrelation} we may
want to write the plasma frequency as a function of momentum $|\bm{k}|$
instead of $\omega$. Since the photon of the scattering process is
free, we can simply write $\omega=|\bm{k}|$ in \eqref{fsa.axion},
\begin{equation}\label{fsa.axion.k}
|f(0)|=\frac{3g^2}{8\pi}\frac{\bm{k}^2m_a}{|4\bm{k}^2-m_a^2|}.
\end{equation}

\subsubsection{Plasma frequency in axion medium}
We obtain from \eqref{fsa.axion} that
\begin{equation}\label{omega_p^2.axion}
\omega_p^2=\frac{3Ng^2}{2}
\frac{\omega^2m_a}{\left|4\omega^2-m_a^2\right|}.
\end{equation}
Recall that our derivation of \eqref{omega_p^2} assumed
$\omega_p^2\ll\omega^2$, and hence \eqref{omega_p^2.axion} is valid when
\begin{equation}
\frac{\frac{3}{2}Ng^2m_a}{\left|4\omega^2-m_a^2\right|}\ll1.
\end{equation}
Furthermore, an energy or momentum dependent plasma frequency must
satisfy the condition \eqref{pf-condition}. When
$\omega>\frac{1}{2}m_a$, the result \eqref{omega_p^2.axion} clearly
satisfies the condition \eqref{pf-condition}, since
$\frac{\partial\omega_p^2}{\partial|\bm{k}|}
=-3Ng^2\frac{|\bm{k}|m_a^3}{(4\bm{k}^2-m_a^2)^2}<0$.

Axions are expected to be very light: $m_a\lesssim10^{-5}\;$eV or
possibly $m_a\lesssim10^{-18}\;$eV \cite{Gorghetto:2021fsn}. For photons
with energies well above the axion mass, $\omega\gg m_a$, the plasma
frequency \eqref{omega_p^2.axion} is nearly constant, i.e. independent
of the frequency of the incoming light:
\begin{equation}\label{hepf.axion}
\omega_p^2=\frac{3}{8}Ng^2m_a\left( 1+\sum_{n=1}^{\infty}
\left(\frac{m_a^2}{4\omega^2}\right)^n \right)\cong\frac{3}{8}Ng^2m_a.
\end{equation}
If the axion is very light $m_a\lesssim10^{-18}\;$eV, the above result
can even be applied to most radio waves.
Notice that the result for the dispersion of light in axion medium given
in \eqref{hepf.axion} using the Green functions method in quantum field
theory, as performed in this work, is new and has not been previously
given in the literature.

In the case of gamma-ray bursts, we are deep in the high energy regime.
Thus the relevant result is
\begin{equation}
\omega_p^2=\frac{3}{8}Ng^2m_a.\label{axionformula}
\end{equation}
Note that \eqref{axionformula} is proportional to the mean mass density
of the axions, $\rho_{a}=Nm_{a}$, as
\begin{equation}
\omega_p^2=\frac{3}{8}g^2\rho_{a}.
\end{equation}
Thus only the density of axionic dark matter determines the dispersion
of light in the cosmic axion medium.

\subsubsection{The delay of gamma-rays in axion medium}
The group velocity of gamma-rays in the axion medium is
\begin{equation}
v_{GRB}= 1-d, \quad d=
\frac{3}{16}\frac{g^2 \rho_{a}}{\omega^2}.
\end{equation}
Estimating the effective coupling constant to be
\begin{equation}
g= 10^{-10}\, \mathrm{GeV}^{-1},
\end{equation}
and assuming, according to the GUT models,
\begin{equation}
g_\gamma = \frac{m_u}{m_d+m_u} \approx 0.36,
\end{equation}
we obtain
\begin{equation}
d=1.4\times 10^{-68}\times\frac{\bar\rho_{a}}{\bar{E}^2},
\end{equation}
where $\bar\rho_{a}$ is the axion density in units
$\mathrm{GeV}/\mathrm{m}^3$ and
$\bar{E}$ is gamma-ray energy in GeV.

Thus the delay of gamma-rays in the axion medium is
\begin{equation}
\tau
=\frac{3}{16}\frac{g^2\rho_{a}}{\omega^2}\times\frac{D}{c}
=1.4\times 10^{-68}\times\frac{\bar\rho_{a}}{\bar{E}^2}
\times\frac{D}{c},\label{delay.axionmed}
\end{equation}
where $D$ the effective distance traveled by the photons. (As mentioned,
we take the velocity of neutrinos to be $c$.)

As a first estimate, let us consider the Galactic Halo (GH) where dark
matter is assumed to consist of axions.
The energy density of GH is $0.45\;\mathrm{GeV}/\mathrm{cm}^3$ and the
radius of GH is $5\times 10^{20}\;\mathrm{m}$ \cite{Braine:2019fqb}.
The delay of GRB propagating through the GH is
\begin{equation}
\tau=\frac{1.1\times 10^{-50}\;\mathrm{s}}{\bar{E}^2}. \label{b}
\end{equation}

As a second example, we consider a massive galaxy filament. The average
axion density is estimated as $\rho_{a}=10^3\;\mathrm{GeV}/\mathrm{m}^3$
and the effective distance is
$D=3\;\mathrm{Gpc}=9\times10^{25}\;\mathrm{m}$. The delay of GRB is
obtained as
\begin{equation}
\tau=\frac{0.7\times 10^{-48}\;\mathrm{s}}{\bar{E}^2}. \label{c}
\end{equation}
The delay is negligible for high-energy photons, as it also is in the
case of GH.

We may also comment on the importance of the magnetic field. For an
electron-positron plasma, there is no magnetic contribution. By extending to the
case of chiral media, the magnetic field comes into play giving an
enhancement of $g$ and a reduction of $N_{a}$. It is known that for a
chiral medium, there is a magnetic contribution to the Casimir effect
\cite{Jiang:2018ivv,Hoye:2020czz}.

\subsection{Derivation of the dispersion relation of light in photon
medium due to axion exchange}
When axions exist, photons can interact with each other via axion
exchange, in addition to the usual photon-by-photon interaction due
to fermion-antifermion loops (box diagrams).

\subsubsection{Calculation of the scattering amplitude}
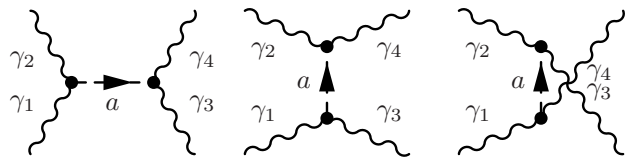
\begin{figure}[ht]
\begin{center}
\begin{fmffile}{photon-diagram}
\begin{fmfgraph*}(27,19)
\fmfleft{i1,i2} \fmfright{o1,o2}
\fmf{photon,label=$\gamma_1$,label.side=left}{i1,v1}
\fmf{photon,label=$\gamma_2$,label.side=right}{i2,v1}
\fmf{scalar,label=$a$}{v1,v2}
\fmf{photon,label=$\gamma_3$,label.side=left}{v2,o1}
\fmf{photon,label=$\gamma_4$,label.side=right}{v2,o2}
\fmfdot{v1,v2}
\end{fmfgraph*}
\begin{fmfgraph*}(27,19)
\fmfleft{i1,i2} \fmfright{o1,o2}
\fmf{photon,label=$\gamma_1$,label.side=left}{i1,v1}
\fmf{photon,label=$\gamma_3$,label.side=left}{v1,o1}
\fmf{scalar,label=$a$}{v1,v2}
\fmf{photon,label=$\gamma_2$,label.side=right}{i2,v2}
\fmf{photon,label=$\gamma_4$,label.side=right}{v2,o2}
\fmfdot{v1,v2}
\end{fmfgraph*}
\begin{fmfgraph*}(27,19)
\fmfleft{i1,i2} \fmfright{o1,o2}
\fmf{photon,label=$\gamma_1$,label.side=left}{i1,v1}
\fmf{phantom}{v1,o1}
\fmf{photon,tension=0,label=$\gamma_4$,label.side=right,label.dist=.5mm}
{v1,o2}
\fmf{scalar,label=$a$}{v1,v2}
\fmf{photon,label=$\gamma_2$,label.side=right}{i2,v2}
\fmf{phantom}{v2,o2}
\fmf{photon,tension=0,label=$\gamma_3$,label.side=left,label.dist=.5mm}
{v2,o1}
\fmfdot{v1,v2}
\end{fmfgraph*}
\end{fmffile}
\end{center}
\caption{Feynman diagrams for scattering of photon on photon in axion
electrodynamics.}
\label{fig1}
\end{figure}

We consider scattering of photon on photon
$\gamma\gamma\rightarrow\gamma\gamma$ at tree level. The scattering
amplitude $\mathcal{M}$ is described by the diagrams in Fig.~\ref{fig1}.
\unitlength = 1mm 

The scattering amplitude is written as
\begin{equation}
\begin{split}
i\mathcal{M}&=\frac{i}{2}g\epsilon^{\alpha\beta\gamma\delta}
k_{3\alpha} \epsilon^{*}_\beta(\bm{k}_3,\lambda_3)k_{4\gamma}
\epsilon^{*}_\delta(\bm{k}_4,\lambda_4) \frac{-i}{s-m_a^2}\\
&\times\frac{i}{2}g\epsilon^{\mu\nu\rho\sigma}k_{1\mu}
\epsilon_{\nu}(\bm{k}_1,\lambda_1) k_{2\rho}
\epsilon_{\sigma}(\bm{k}_2,\lambda_2)\\
&+\frac{i}{2}g\epsilon^{\alpha\beta\gamma\delta}
k_{2\alpha} \epsilon_\beta(\bm{k}_2,\lambda_2)k_{4\gamma}
\epsilon^{*}_\delta(\bm{k}_4,\lambda_4) \frac{-i}{t-m_a^2}\\
&\times\frac{i}{2}g\epsilon^{\mu\nu\rho\sigma}k_{1\mu}
\epsilon_{\nu}(\bm{k}_1,\lambda_1) k_{3\rho}
\epsilon^{*}_{\sigma}(\bm{k}_3,\lambda_3)\\
&+\frac{i}{2}g\epsilon^{\alpha\beta\gamma\delta}
k_{2\alpha} \epsilon_\beta(\bm{k}_2,\lambda_2)k_{3\gamma}
\epsilon^{*}_\delta(\bm{k}_3,\lambda_3) \frac{-i}{u-m_a^2}\\
&\times\frac{i}{2}g\epsilon^{\mu\nu\rho\sigma}k_{1\mu}
\epsilon_{\nu}(\bm{k}_1,\lambda_1) k_{4\rho}
\epsilon^{*}_{\sigma}(\bm{k}_4,\lambda_4).
\end{split}\label{M.photon-photon}
\end{equation}
where $s=(k_1+k_2)^2=2k_1\cdot k_2$, $t=(k_1-k_3)^2=-2k_1\cdot k_3$ and
$u=(k_1-k_4)^2=-2k_1\cdot k_4$.

We use the center-of-momentum frame, where $k_1=(\omega,\bm{k})$,
$k_2=(\omega,-\bm{k})$, $k_3=(\omega',\bm{k}')$ and
$k_4=(\omega',-\bm{k}')$. We have $s=2(\omega^2+\bm{k}^2)=4\omega^2$,
$t=-2(\omega\omega'-\bm{k}\cdot\bm{k}')=-2\omega\omega'(1-\cos\theta)$
and
$u=-2(\omega\omega'+\bm{k}\cdot\bm{k}')=-2\omega\omega'(1+\cos\theta)$,
where $\theta$ is the scattering angle between $\bm{k}$ and $\bm{k}'$.
Conservation of energy and momentum ensures that $s+t+u=0$, and
implies that $\omega'=\omega$.

The differential cross section in the center-of-momentum frame for
unpolarized photons is
\begin{equation}
\frac{d\sigma}{d\Omega}=\frac{1}{64\pi^2s}\left(
\frac{1}{4}\sum_{\lambda_1,\lambda_2,\lambda_3,\lambda_4}
|\mathcal{M}|^2 \right).\label{dcs.cm}
\end{equation}

We consider scattering at angle $\theta=0$. Then the four-momenta
$k_3=k_1$ and $k_4=k_2$, and the same polarization vectors are used to
describe the corresponding initial and final photons that have the same
momenta. In Cartesian coordinates, we may choose
$k_1=k_3=(\omega,0,0,\omega)$ and linear polarization vectors
$\epsilon(\bm{k}_1,1)=\epsilon(\bm{k}_3,1)=(0,1,0,0)$ and
$\epsilon(\bm{k}_1,2)=\epsilon(\bm{k}_3,2)=(0,0,1,0)$. Correspondingly,
we have $k_2=k_4=(\omega,0,0,-\omega)$ and polarization vectors
$\epsilon(\bm{k}_2,1)=\epsilon(\bm{k}_4,1)=(0,1,0,0)$ and
$\epsilon(\bm{k}_2,2)=\epsilon(\bm{k}_4,2)=(0,0,-1,0)$.
Then we need to evaluate the vertices of the amplitude
\eqref{M.photon-photon}.

First consider the vertices in the $s$-channel contribution. In the
vertices we have
\begin{equation}
\epsilon_{\mu\nu\rho\sigma}k_{1}^{\mu}\epsilon^{\nu}(\bm{k}_1,\lambda_1)
k_{2}^{\rho}\epsilon^{\sigma}(\bm{k}_2,\lambda_2)=-2\omega^2
J_{\lambda_1\lambda_2},
\end{equation}
and
\begin{equation}
\epsilon_{\mu\nu\rho\sigma}k_{3}^{\mu}\epsilon^{*\nu}(\bm{k}_3,
\lambda_3)k_{4}^{\rho}\epsilon^{*\sigma}(\bm{k}_4,\lambda_4)=-2\omega^2
J_{\lambda_3\lambda_4},
\end{equation}
where we introduced the two-dimensional anti-diagonal unit matrix
$J=\left(\begin{matrix}0&1\\ 1&0\end{matrix}\right)$.

In the $t$-channel contribution, the two momenta involved in each
vertex are identical, and consequently the contribution vanishes:
\begin{multline}
\epsilon_{\mu\nu\rho\sigma}k_{1}^{\mu}\epsilon^{\nu}(\bm{k}_1,\lambda_1)
k_{3}^{\rho}\epsilon^{*\sigma}(\bm{k}_3,\lambda_3)\\
=\epsilon_{\mu\nu\rho\sigma}k_{1}^{\mu}\epsilon^{\nu}(\bm{k}_1,
\lambda_1)k_{1}^{\rho}\epsilon^{*\sigma}(\bm{k}_1,\lambda_3)=0,
\end{multline}
and
\begin{multline}
\epsilon_{\mu\nu\rho\sigma}k_{2}^{\mu}\epsilon^{\nu}(\bm{k}_2,\lambda_2)
k_{4}^{\rho}\epsilon^{*\sigma}(\bm{k}_4,\lambda_4)\\
=\epsilon_{\mu\nu\rho\sigma}k_{2}^{\mu}\epsilon^{\nu}(\bm{k}_2,
\lambda_2)k_{2}^{\rho}\epsilon^{*\sigma}(\bm{k}_2,\lambda_4)=0.
\end{multline}

In the $u$-channel contribution, we obtain
\begin{equation}
\epsilon_{\mu\nu\rho\sigma}k_{1}^{\mu}\epsilon^{\nu}(\bm{k}_1,\lambda_1)
k_{4}^{\rho}\epsilon^{*\sigma}(\bm{k}_4,\lambda_4)=-2\omega^2
J_{\lambda_1\lambda_4},
\end{equation}
and
\begin{equation}
\epsilon_{\mu\nu\rho\sigma}k_{2}^{\mu}\epsilon^{\nu}(\bm{k}_2,
\lambda_2)k_{3}^{\rho}\epsilon^{*\sigma}(\bm{k}_3,\lambda_3)=-2\omega^2
J_{\lambda_2\lambda_3}.
\end{equation}

The unpolarized squared amplitude for the scattering angle $\theta=0$ is
\begin{multline}
\frac{1}{4}\sum_{\lambda_1,\lambda_2,\lambda_3,\lambda_4}
|\mathcal{M}(0)|^2=\frac{g^4\omega^8}{(s-m_a^2)^2}\\
+\frac{g^4\omega^8}{2(s-m_a^2)(u-m_a^2)}
+\frac{g^4\omega^8}{(u-m_a^2)^2}.
\end{multline}
For angle $\theta=0$ we have $s=u=4\omega^2$ and hence we obtain
\begin{equation}
\frac{1}{4}\sum_{\lambda_1,\lambda_2,\lambda_3,\lambda_4}
|\mathcal{M}(0)|^2
=\frac{5g^4\omega^8}{2(4\omega^2-m_a^2)^2}.
\end{equation}
Thus the differential cross section \eqref{dcs.cm} for unpolarized
photons at the scattering angle $\theta=0$ is
\begin{equation}
\frac{d\sigma(0)}{d\Omega}=\frac{5g^4}{2^{9}\pi^2}
\frac{\omega^6}{(4\omega^2-m_a^2)^2}.
\end{equation}

\subsubsection{Dispersion relation and the delay of gamma-rays}
The contribution of axions to the photon plasma frequency is
\begin{equation}
\omega_p^2=\frac{\sqrt{5}Ng^2}{4\sqrt{2}}
\sqrt{\frac{\omega^6}{(4\omega^2-m_a^2)^2}}
=\frac{\sqrt{5}Ng^2}{16\sqrt{2}}
\frac{\omega}{1-\frac{m_a^2}{4\omega^2}},
\end{equation}
where $N$ is the number density of photons.
In high energies $\omega\gg m_a^2$, we have
\begin{equation}
\omega_p^2=\frac{\sqrt{5}Ng^2}{16\sqrt{2}}\omega\left(
1+\sum_{i=1}^\infty \left(\frac{m_a^2}{s}\right)^i \right)
\simeq \frac{\sqrt{5}Ng^2}{16\sqrt{2}}\omega,\label{pf.photon.a}
\end{equation}
where $\omega=\sqrt{s}/2$ in terms of the invariant $s$.

The group velocity of gamma-rays in the photon medium is
\begin{equation}
v_{GRB}= 1-d, \quad d=\frac{\sqrt{5}Ng^2}{32\sqrt{2}\omega}.
\end{equation}
Using the photon number density of CMB,
$N_{\gamma}=(4$--$5)\times10^{8}\;\mathrm{m}^{-3}$,
and estimating the effective coupling constant to be
$g= 10^{-10}\,\mathrm{GeV}^{-1}$,
the contribution of axion exchange to the GRB delay produced by CMB is
given as
\begin{equation}
\tau_{\mathrm{CMB}}
=\frac{3\times10^{-54}}{\sqrt{\bar{E}}}\times\frac{D}{c}.
\end{equation}

For GRB originating from the farthest galaxies
($D=1.4\times10^{26}\;\mathrm{m}$) the delay produced by
the interaction with CMB is
$\tau_{\mathrm{CMB}}=2\times10^{-36}\;\mathrm{s}/\sqrt{\bar{E}}$;
e.g. for gamma-ray energy $E=100\;\mathrm{GeV}$ the delay is
$\tau_{\mathrm{CMB}}=2\times10^{-37}\;\mathrm{s}$,
which is a negligible delay.

\section{Remarks on Lorentz invariance violation}
\label{secLIV}
The high energy tests of (an intrinsic) Lorentz invariance violation (LIV), as
proposed by \cite{Coleman:1998ti} with specific examples for them,
have attracted considerable interest in connection with GRB; see e.g. \cite{AmelinoCamelia:1997gz,Ellis:1999sd,Jacob:2006gn,Jacob:2008bw,AmelinoCamelia:1996pj,RodriguezMartinez:2006ee,Xiao:2009xe,Shao:2009bv,Zhang:2014wpb,Li:2021tcw}.
We shall now compare the present light dispersion approach to the
assumption of Lorentz invariance violation. Typically, in the LIV
approach motivated by quantum gravity effects the dispersion relation
contains higher-powers of energy,
\begin{equation}
E^2\left[ 1+\xi\frac{E}{E_{QG}}+O\left(\frac{E^2}{E_{QG}^2}\right)
\right]=\bm{p}^2c^2+m^2c^4,
\end{equation}
where $E_{QG}$ is an effective energy scale for quantum gravity,
commonly taken to be of order $10^{16}\;\mathrm{GeV}$, and $\xi$ is an
arbitrary parameter. The expression for the group velocity takes the
following form to leading order in $E/E_{QG}$,
\begin{equation}
v_g= c\left( 1-\xi \frac{E}{E_{QG}}\right). \label{LIV}
\end{equation}
Then it is  assumed that $\xi>0$ so that $v_g$ as the signal speed of
radiation is subluminal. Thus, the higher energy of GRB, the greater
slowdown of it.

Comparison with the dispersion relation \eqref{groupvelocity} above
leads to an interesting observation: the energy dependencies of the
resulting GRB time delays are qualitatively different, which can be used
to distinguish the two interpretations including their amounts
experimentally. In the case of LIV, the time delay behaves as $\tau\propto E$. Dispersion in electron and axion media produces a time delay as $\tau\propto E^{-2}$.
Thus, if gamma-rays with the highest energies arrive first, the conventional
dispersive plasma theory with an electron-like coupling is supported,
while if the highest-energy gamma-rays arrive later, the Lorentz
invariance breaking dispersion relation is supported. It would be quite
important to test out this issue experimentally. An observation of
advance of the highest energy photons would imply that the LIV model
would have to be reconsidered. In order to perform such a test, the
spectrum of GRB needs to be recorded with as high temporal resolution as
possible.

Since the LIV modification of the dispersion relation should effect
particles of all kinds, and particularly both photons and neutrinos, it
is difficult to explain the delays between GRB and neutrino bursts
coming from the same source. In this interpretation such a delay could
only be possible if the energy of neutrinos differs from the energy of
photons by several orders of magnitude \cite{Jacob:2006gn}, which is
hard to believe. However, if one would argue that in LIV the dispersions
relations for photon and neutrino are different  due to  different
quantum gravity effects on them, then by all means the whole
effect should not be called LIV but the \emph{induced} LIV.

It is noteworthy that it is not known experimentally whether
the gamma-ray bursts arrive earlier, or later, than the neutrino bursts
--- the IceCube in Antarctica has detected some cosmic neutrinos, but
they cannot be associated with any astrophysical object \cite{rodrigo}
(for a recent experiment at this detector, see \cite{Aartsen:2020eof}).

Finally, we would like to mention that an \emph{intrinsic} violation
of the Lorentz symmetry can be detected in several ways in precision
experiments in different processes as has been in full details described
in \cite{Coleman:1998ti}. LIV can also occur  in a way that differs from
the hitherto used approach  to LIV by changing the Special Relativity
dispersion relation. For instance, the Lorentz group could be broken
to  certain proper subgroups of the Poincar\'e group, to the so-called
Very Special Relativity \cite{Cohen:2006ky}. It also can occur that the
dispersion relation does not change but the Lorentz symmetry is broken
--- an example is the noncommutative field theory
\cite{Ardalan:1998ce,Seiberg:1999vs}, where we have the residual
twisted Poincar\'e symmetry \cite{Chaichian:2004za}.

\section{Discussion and Conclusions}

We have considered the propagation of gamma-ray photons in the
interstellar space --- a problem of considerable interest by itself
and analyzed in details the dispersion of light travelling through the
interstellar space behaving as a plasma medium.

Dispersive properties of normal matter and background radiation, CMB,
are insufficient to produce a significant GRB delay, when using some
indicative densities of the constituent particles.

We have considered the media having nonrelativistic constituents -- the nonrelativistic approximation can be justified since the main part of the GRB travel occurs in the interstellar space, where the delay actually occurs, and the temperature is so low (as in CMB) that the constituent particles of the media are practically frozen.

Then we have taken into account the effect of dark matter by assuming
that dark matter consists of very light and cold particles, the
hypothetical axions, with their theoretically expected coupling to photons.
We find that for the hitherto realistic density of dark matter (revealed
by its gravitational effect), the produced GRB delay is also very small.

The derived basic formulas for the time delay $\tau$ are given in
Eq.~(\ref{delay.ep}) for an electron-positron plasma, in Eq.~(\ref{d}) for a
photon plasma as CMB medium, in Eq.~(\ref{b}) for a Galactic Halo
and in (\ref{c}) for a massive galaxy filament. The numerical values
in those Eqs. are for the realistic values of the particle densities
according to different estimates, or the ones considered to be more
or less reasonable. The derivations of the light dispersion relation in
all the different media in QFT are presented in the
Appendix~\ref{appendix} and its subsections.

As seen from the Eqs.~(\ref{delay.ep}), (\ref{d}), (\ref{b}) and (\ref{c}),
the smallest delay for the high energy gamma-rays comes from
dispersion in an axion plasma, and slightly larger delay for an electron-positron
plasma.

The largest delay comes from the dispersion of light in a medium
such as filled with CMB. In addition, the delay time as a function of the
incoming photon energy in a medium filled with light radiation is $1/E^{3/2}$,
while in an electron or axion medium it is $1/E^2$.
Thus, at higher and higher energies the dispersion in light medium
dominates over the other two media.
The latter, is a quite important issue, since the value for the delay,
which is proportional to the density of particles, given in Eq.~(\ref{d})
is estimated with the average density in CMB in only the observed part
of the Universe, (being typically $\tau\sim2\times10^{-8}\;\mathrm{s}$ for a
100 GeV gamma-ray burst from farthest galaxies), while in other parts
of the Universe the density of photons can be by far larger.

As already mentioned, we  have considered the dispersion for each medium separately, in order to evaluate and find out which medium gives the largest delay for  the dispersion of light, which comes out to be the CMB medium. Considering the total dispersion due to  all the three media, gives simply the addition of each of them. This is easily seen from the derivation of dispersion relations in QFT and the fact that the interference terms among the scattering amplitudes with different final states do not give a contribution in the square of the absolute value of the sum in obtaining the cross section and thus the sum will be the addition of the absolute magnitudes of the diagrams with the same final states.

The very high energies and long travel distances of GRB have initiated
to consider different models of Lorentz violation by modifying the
(mass-energy-momentum) dispersion relations. Such an approach typically
leads to a signal velocity as in Eq.~\eqref{LIV}, which decreases as a
function of energy. We also see that such breaking of Special Relativity
would lead to the simultaneous breaking of General Relativity, in which
case many other results should be reconsidered and revalued.
Thus, we believe that before invoking the drastic assumption of
breaking of the Special Relativity to interpret GRB delays, one should
also consider all possible ways to explain the phenomenon within the
standard physics, a special case of which, namely the dispersion of
light in the interstellar space, what always exists, has been presented
in the present work.

As always, the decisive evidence will finally be given by experiment:
in the case of the interpretation in terms of dispersion of light, the
higher-energy gamma-rays arrive the Earth earlier and the lower-energy
ones later, while in the typical LIV models hitherto presented based on
quantum gravity effects such as in stringy foam models for the space,
it occurs the opposite.

We should emphasize that all the models considered till now can only be
called as an \emph{induced (seeming)} LIV and not simply as LIV,
which could imply \emph{intrinsic (genuine)} LIV. The wording of simply
LIV has been used for brevity in some recent literature instead of the
\emph{induced} LIV.

Concerning the experimental observation of an \emph{intrinsic}
(genuine) violation of the Lorentz invariance and its theoretical
interpretation, we have already mentioned at the very end of
Sect.~\ref{secLIV} above, and referred to the seminal works in
\cite{Coleman:1998ti} and \cite{Cohen:2006ky}. But we would like to
recall that all the symmetries and laws derived from them are based
on the existence of some group of symmetries, no matter whether they
are global or local (gauge), internal or spacetime (external) symmetries and
their breaking described by a residual or subgroup of the original
groups of symmetries or an acceptable deformation of those groups,
i.e. ones with their motivation not based on the presence of a medium such as gravity or curvature, but not breaking the Lorentz group in an arbitrary way.
The same should be also in the case of the breaking of the Lorentz symmetry.
The work in \cite{Cohen:2006ky} is a good example, where the natural
requirements for the residual group of symmetry, i.e. the broken ones,
have been chosen to be a subgroup of the Poincar\'e group, and such that
both the translational and CPT invariance remain preserved.

Several experiments, e.g. the recent
Refs.~\cite{Albert:2019nnn,Acciari:2019dbx,Acciari:2020kpi}, will
analyze the energy-dependent delay in the arriving time of photons and
will be able to resolve this issue. In the dispersive approach, the delay
of an electromagnetic signal increases at lower photon energies.
Therefore signals with radio frequencies would be a possible way
to test the approach. Since the mass of axion is very low, possibly
as low as $m_a\sim10^{-18}\;\mathrm{eV}$, and our derivations
hold for photons with energy much higher than the axion mass,
our results may be applicable to many radio signals as well.

The observation of gravitational waves (GW) and a short GRB from the
merger of two neutron stars in NGC 4993
\cite{TheLIGOScientific:2017qsa,GBM:2017lvd} shows that gamma-ray
photons do not experience long delays, since the GRB was observed
only 1.7 seconds after the GW. Electromagnetic signals with lower
frequencies were recorded by several teams starting eleven hours after
the GW and ending weeks later, ranging from x-rays to radio frequencies
\cite{GBM:2017lvd}. Optical and infrared observations showed a
towards-red evolution during 10 days. That is consistent with emissions from
a cooling debris of the merger. We remark two things. Firstly, the GRB
delays caused by dispersion of gamma-rays in the cosmic medium,
as derived in this work, are much shorter than the observed delay
between GW 170817 and GRB 170817A signals, assuming that the GW
are not effected by the medium. That is expected, since
the gamma-rays are thought to be produced after the gravitational waves.
Secondly, for the dispersive approach the dependency of the time delay on
photon frequency (i.e. higher frequencies are delayed less than lower
frequencies) is consistent with the observations of the merger. The same
might not be the case with LIV models which exhibit an opposite relation
between the time delay and the photon frequency.

In our estimates we have considered only the average axion densities
at great distances, particularly at the level of galaxy filaments and
in the dark matter halo of a typical galaxy.
Any finer details of axion distribution have been neglected in these
estimates. The same concerns the density of photons, such as in CMB,
taken to be as in only the observed part of the Universe.
Therefore, it would be enlightening to study the effects of different
axion distributions, as well as the densities of the photons in the
larger distances in greater detail. In this way, future studies of
GRB delays produced by dispersion of light in media can also
shed additional light on the microstructure of the Universe.

As some final remarks, we would like to emphasize the following points:\\
i) in the case of an \emph{intrinsic} LIV, i.e. a \emph{genuine} violation of the Special Relativity, when the usual dispersion relation between the energy, momentum and mass of a particle is changed, that means the Minkowski metric is changed. Consequently, the General Relativity, based on the generalisation of   Minkowski metric/diffeomorphism invariance, will break too and its usual consequences cannot anymore be used for analysing the experimental data;\\
ii) in all the models so far proposed concerning the GRB with the arguments based on quantum gravity effects such as stringy foam, or connected to the curvature of the space and interpreting them as LIV, they are actually induced LIV. Although calling them LIV is intriguing, that is not correct;\\
iii) the dispersion relation being simply a kinematical equation, is \emph{universal} and depends on the metric of the spacetime and not on the nature of a particle or on any moving body. Therefore, the above-mentioned kind of models for LIV cannot explain any delay of high energy GRB compared with neutrino bursts,
since the tiny masses of neutrinos have no effect on the group velocity, unless
one tries to invoke different interactions of the media with GRB and neutrino --- per se they are \emph{induced Lorentz invariance violations}.

\vspace{1em}
\noindent
\underline{Note added:}

\vspace{0.3em}
\noindent
After the completion of our work we were informed about two previous works
\cite{Balakin:2013qya} and \cite{Dobrynina:2014qba} on the propagation of light in an axion or photon plasmas, respectively, in quite different aspects and using different methods of calculation but without addressing the issue of GRB delay.

The work in \cite{Balakin:2013qya} gives a comprehensive account of dispersion of light in a relativistic plasma theory, including also some numerical evaluation of the results. Their model is somewhat different from ours, since it involves nonminimal gravitational interaction for the pseudoscalar (axion) field, and the axion field varying linearly with respect to time.
A main objective of their work was to show that the phase velocity of transverse electromagnetic waves can also be less than $c$, providing the possibility of resonant plasma/waves interactions.
Within the permitted space of parameters for the propagating electromagnetic waves, the squared plasma frequency, $\omega_p^2$, and consequently the effective photon mass squared, $m^2_\gamma$, are positive (see Fig.~3).

The work \cite{Dobrynina:2014qba} has considered the conversion of high-energy gamma-rays to axions and a dominant dispersion of light in a photon medium corresponding to CMB effect. Using the nonlinear Euler-Heisenberg electrodynamics Lagrangian they obtain a negative value for the squared effective photon mass, $m^2_\gamma<0$. That would correspond to a superluminal group velocity for GRB. We believe that this result is incorrect, possibly due to obscure way of calculating the plasma frequency $\omega_p^2$ and obtaining a negative value for it. Using the same way of calculation for an axion medium, they would also obtain a negative value of $\omega_p^2$.

\subsection*{Acknowledgments}
We are much grateful to Felix Aharonian, Stanley Deser, Merab Gogberashvili,
Rodrigo Gracia-Ruiz, Friedrich Hehl, Archil Kobakhidze, Bo-Qiang Ma,
Claus Montonen, Viatcheslav Mukhanov, Yuri Obukhov, Adam Schwimmer,
Anca Tureanu and Grigory Volovik for many useful discussions at different stages of the work.

\appendix
\numberwithin{equation}{section}
\section{Appendix: Calculation of the plasma frequency from quantum
field theory}\label{appendix}
Here we explain how the refraction index and the plasma frequency can be
derived from quantum field theory. In order to show how it works, we
first consider the case of electrons and photons with standard quantum
electrodynamics. Then we consider the axion with its interaction to the
electromagnetic field \eqref{L_int}.
We use a system of units with Heaviside--Lorentz electromagnetic units
and $\hbar=c=1$.

\subsection{Preliminaries on optical theorem}
The refraction index is related to the forward scattering amplitude $f(0)$
(at zero scattering angle) as \cite{Jauch-Rohrlich,Newton}
\begin{equation}\label{n.qm}
n=1+\frac{2\pi N f(0)}{\omega^2},
\end{equation}
where $N$ is the number density of scatterers, that is the density
of the constituents of the plasma. The relation \eqref{n.qm}
is valid when $n$ is close to one, $|n-1|\ll1$, and follows from the
inteference between incident and scattered waves.
In \eqref{n.qm}, $n$ is a complex number, where the real part describes dispersion and the imaginary part describes absorption.
Together with the relation $\Imag n=2\pi\,N\Imag f(0)/\omega^2=N\sigma/2\pi$
between the imaginary part of $n$ and the absorption coefficient
$N\sigma$, where $\sigma$ is the total cross section, \eqref{n.qm} leads
to the Bohr--Peierls--Placzek relation \cite{Bohr-Peierls-Placzek} (see
also \cite{Jauch-Rohrlich,Newton}),
\begin{equation}\label{opticaltheorem}
\sigma=\frac{4\pi\Imag f(0)}{\omega},
\end{equation}
also known as the optical theorem.

We notice that within the perturbative quantum field theory the neglect
of $\Imag f(0)$ is naturally satisfied: one can see from \eqref{opticaltheorem}
that the imaginary part of $f(0)$ is proportional to the cross section,
which is proportional to the square of the amplitude and thus higher
in the order of the small coupling constant than in the corresponding
formula in \eqref{fsa.abs}, which is of the lower order in the coupling constant,
being proportional to the amplitude itself.

Inserting \eqref{refractionindex} into \eqref{n.qm} we obtain a
relation between the plasma frequency and the scattering amplitude,
\begin{equation}
\sqrt{1-\frac{\omega_p^2}{\omega^2}}=1+\frac{2\pi N\Real f(0)}{\omega^2}.
\end{equation}
When the photon frequency is large compared to the plasma frequency,
$\omega^2\gg\omega_p^2$, we obtain
\begin{equation}\label{omega_p^2}
\omega_p^2=-4\pi N \Real f(0).
\end{equation}

The scattering amplitude $f(\theta)$ is defined as a part of the
quantum mechanical wave function at large distance $r$ from the
scatterer,
\begin{equation}
\psi(\bm{r})=C\left( e^{i\bm{k}\cdot\bm{r}}+f(\theta)\frac{e^{ikr}}{r}
\right),
\end{equation}
where $C$ is a normalization factor. The differential cross section is
given in terms of the scattering amplitude as
\begin{equation}\label{dsigma}
d\sigma(\theta)=|f(\theta)|^2d\Omega.
\end{equation}
The closest thing to the scattering amplitude $f$ in quantum field
theory is the so-called $T$-matrix defined as the scattering part of the
$S$-matrix
\begin{equation}
\braket{\psi_f|S|\psi_i}=\braket{\psi_f|\psi_i}
+\braket{\psi_f|iT|\psi_i}.
\end{equation}
The $T$-matrix, i.e. the Green's function, is related to the Feynman
invariant scattering amplitude $\mathcal{M}$.
In order to compute $f(0)$ in quantum field theory, we shall use the
differential cross section.
We calculate in quantum field theory the differential cross section $d\sigma(\theta)$
and from \eqref{dsigma} the forward scattering amplitude follows as
\begin{equation}\label{fsa.abs}
|f(0)|=\left( \frac{d\sigma(0)}{d\Omega} \right)^{\frac{1}{2}}.
\end{equation}

After the preliminaries presented above, we shall now proceed with
the derivation of the plasma frequency for electron-positron plasma in standard
quantum electrodynamics, as well as for a photon medium.
Calculation of the plasma frequency for an electron-positron plasma is given here
in order to show that our results coincide with the classical ones which
have been derived before from the classical Maxwell equations.

\subsection{Electron-positron plasma}
\subsubsection{Scattering amplitude in quantum electrodynamics}
\begin{figure}[ht]
\begin{center}
\begin{fmffile}{electron-diagrams}
\begin{fmfgraph*}(34,22)
\fmfleft{i1,i2} \fmfright{o1,o2}
\fmf{electron,label=$e^{-}(p,,s)$,label.side=left}{i1,v1}
\fmf{photon,label=$\gamma(k,,\lambda)$,label.side=right}{i2,v1}
\fmf{electron,label=$e^{-}$}{v1,v2}
\fmf{electron,label=$e^{-}(p',,s')$,label.side=left}{v2,o1}
\fmf{photon,label=$\gamma(k',,\lambda')$,label.side=right}{v2,o2}
\fmfdot{v1,v2}
\end{fmfgraph*}
\hspace{5mm}
\begin{fmfgraph*}(34,22)
\fmfleft{i1,i2} \fmfright{o1,o2}
\fmf{electron,label=$e^{-}(p,,s)$,label.side=left}{i1,v2}
\fmf{photon,label=$\gamma(k,,\lambda)$,label.side=right,label.dist=7}{i2,v1}
\fmf{electron,label=$e^{-}$}{v2,v1}
\fmf{photon,label=$\gamma(k',,\lambda')$,label.side=left,label.dist=7}{v2,o1}
\fmf{electron,label=$e^{-}(p',,s')$,label.side=right,label.dist=7}{v1,o2}
\fmfdot{v1,v2}
\end{fmfgraph*}
\end{fmffile}
\end{center}
\caption{Feynman diagrams for scattering of photon on electron.}
\label{fig.e}
\end{figure}
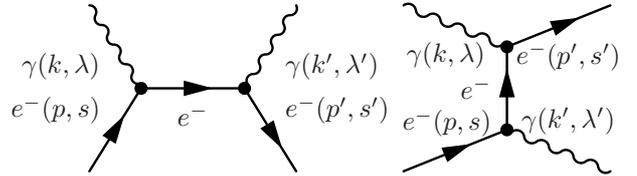

We consider photon-electron scattering
$\gamma+e^{-}\rightarrow\gamma+e^{-}$ at tree level. The scattering
amplitude $\mathcal{M}$ is a sum of the two diagrams in
Fig.~\ref{fig.e}.
\unitlength = 1mm 

The incoming and outgoing electrons are represented by the Dirac spinors
 $u(\bm{p},s)$ and $u(\bm{p}',s')$, respectively. The incoming and
outgoing photons are represented by the polarization vectors
$\epsilon_{\mu}(\bm{k},\lambda)$ and
$\epsilon^{*}_{\mu}(\bm{k}',\lambda')$, respectively.
The amplitude in Feynman gauge is
\begin{equation}
\begin{split}
i\mathcal{M}&=\bar{u}(\bm{p}',s')(-ie\gamma^{\mu})\epsilon^{*}_{\mu}
(\bm{k}',\lambda')\frac{i(\slashed{p}+\slashed{k}+m_e)}{(p+k)^2-m_e^2}\\
&\times\epsilon_{\nu}(\bm{k},\lambda)(-ie\gamma^{\nu})u(\bm{p},s)\\
&+\bar{u}(\bm{p}',s')(-ie\gamma^{\nu})\epsilon_{\nu}
(\bm{k},\lambda)\frac{i(\slashed{p}-\slashed{k}'+m_e)}{(p-k')^2-m_e^2}\\
&\times\epsilon^{*}_{\mu}(\bm{k}',\lambda')(-ie\gamma^{\mu})u(\bm{p},s)\\
&=-ie^2\epsilon^{*}_{\mu}(\bm{k}',\lambda')
\epsilon_{\nu}(\bm{k},\lambda)\bar{u}(\bm{p}',s')\\
&\times\biggl( \frac{\gamma^{\mu}\slashed{k}\gamma^{\nu}+2\gamma^{\mu}p^{\nu}}
{2p\cdot k}
+\frac{\gamma^{\mu}\slashed{k}'\gamma^{\nu}
-2\gamma^{\nu}p^{\mu}}{2p\cdot k'} \biggr)u(\bm{p},s),
\end{split}
\end{equation}
where in the second equality we have written $p^2=m_e^2$, $k^2=0$ and
$(\slashed{p}+m_e)\gamma^{\nu}u(\bm{p},s)=2p^{\nu}u(\bm{p},s)$.

The photons and electron are not polarized, and hence we average over
initial spin states and sum over final spin states. In the squared
Feynman amplitude, we use the spinor completeness relations and perform
the resulting four traces of the $\gamma$-matrices. The result is
\begin{multline}\label{M2_egamma}
\frac{1}{4}\sum_{s,\lambda,s',\lambda'}|\mathcal{M}|^2
=2e^4\Biggl[ \frac{p\cdot k'}{p\cdot k}+\frac{p\cdot k}{p\cdot k'}\\
+2m_e^2\left( \frac{1}{p\cdot k}-\frac{1}{p\cdot k'} \right)
+m_e^4\left( \frac{1}{p\cdot k}-\frac{1}{p\cdot k'} \right)^2 \Biggr].
\end{multline}
In order to obtain the forward scattering amplitude $f(0)$, we obtain
the differential cross section and compare to \eqref{fsa.abs}.
The differential cross section is
\begin{equation}
d\sigma=\frac{1}{2k_0}\frac{1}{2p_0}
\left( \frac{1}{4}\sum_{s,\lambda,s',\lambda'}|\mathcal{M}|^2 \right)
d\mathrm{Lips},
\end{equation}
where the relative velocity of the initial particles is $c=1$ (in any
frame) and the Lorentz invariant two-body phase space is defined as
\begin{equation}\label{dLips}
d\mathrm{Lips}=(2\pi)^4\delta^4(k'+p'-k-p)
\frac{d^3\bm{k}'}{(2\pi)^32k'_0}\frac{d^3\bm{p}'}{(2\pi)^32p'_0}.
\end{equation}
Integration over $\bm{p}'$ is trivial and gives
\begin{equation}
d\mathrm{Lips}=\frac{1}{(2\pi)^24k'_0p'_0}\delta(k_0+p_0-k'_0-p'_0)
d^3\bm{k}',
\end{equation}
where $d^3\bm{k}'=\bm{k}^{\prime 2}d|\bm{k}'|d\Omega$. Since we
consider a single scattering process in vacuum, the photon energy and
momentum are related by the vacuum dispersion relation $k'_0=|\bm{k}'|$.
Integration over $|\bm{k}'|$ gives
\begin{equation}\label{dLips.2}
d\mathrm{Lips}=\frac{(k_0+p_0-p'_0)^2}{(2\pi)^24k'_0p'_0}
d\Omega=\frac{k'_0}{(2\pi)^24p'_0}d\Omega.
\end{equation}

We now choose the rest frame of the initial electron, $p=(m_e,\bm{0})$,
and specialize to the case when the momenta of the initial and final
photons are parallel. When $\bm{k}$ and $\bm{k}'$ are parallel, we have
$k\cdot k'=k_0k'_0-\bm{k}\cdot\bm{k}'
=|\bm{k}||\bm{k}'|(1-\cos\theta)=0$, where $\theta=0$ is the angle
between $\bm{k}$ and $\bm{k}'$. We obtain from conservation of
four-momentum that $(p+k-k')^2=p^{\prime 2}$, where the left-hand side
is $m_e^2+2p\cdot k-2p\cdot k'=m_e^2+2m_e(k_0-k'_0)$
and the right-hand side is $m_e^2$. This implies $k'_0=k_0$. Since
energy is conserved, we also have $p'_0=p_0$. For $\theta=0$ the
squared Feynman amplitude \eqref{M2_egamma} becomes
\begin{equation}\label{sumM2.0}
\frac{1}{4}\sum_{s,\lambda,s',\lambda'}|\mathcal{M}(0)|^2
=4e^4.
\end{equation}
Hence the differential cross section for $\theta=0$ is
\begin{equation}
\begin{split}
d\sigma(0)&=\frac{1}{64\pi^2m_e^2}
\left( \frac{1}{4}\sum_{s,\lambda,s',\lambda'}|\mathcal{M}(0)|^2 \right)
d\Omega\\
&=\frac{e^4}{16\pi^2m_e^2}d\Omega.
\end{split}
\end{equation}
Comparing to \eqref{fsa.abs} gives the absolute value of the forward
scattering amplitude as
\begin{equation}\label{fsa_egamma}
|f(0)|=\frac{\alpha}{m_e},
\end{equation}
where $\alpha$ is the fine-structure constant,
$\alpha=\frac{e^2}{4\pi}$.

\subsubsection{Electron-positron plasma frequency}
The plasma frequency is obtained from \eqref{omega_p^2} and
\eqref{fsa_egamma} as
\begin{equation}
\omega_p^2=\frac{4\pi\alpha N_e}{m_e}=\frac{N_ee^2}{m_e}.
\end{equation}
This is the same result that is obtained from classical electrodynamics
\cite{Langmuir,Jackson,CMRT-ED}.
The electron-positron plasma frequency has also been obtained from quantum
statistical physics by considering interaction in an electron-ion plasma
\cite{Abrikosov}.

\subsection{Photon medium}\label{appendix.photon}
The differential cross section of photon-photon scattering for
high energy $\omega\gg m_e$ and the angle $\theta$ close to zero is
given in the center-of-momentum frame as \cite{Berestetsky:1982}
\begin{equation}
d\sigma=\frac{\alpha^4}{\pi^2\omega^2}\log^4\left(\frac{\omega}{m_e}
\right) d\Omega ,\quad \theta\ll\frac{m_e}{\omega},
\end{equation}
where $\omega=\sqrt{s}/2$.

The observations of GRB are performed in a frame where the GRB photons
have very high energies, while the CMB photons have much lower energies.
In such a frame, the average of $s$ can be obtained by averaging over
the angle between the momenta of the incoming photons for the scattering
of GRB photons on CMB photons:
$\langle s\rangle=\frac{1}{\pi}\int_0^\pi 2\omega_1\omega_2
(1-\cos\phi)d\phi=2\omega_1\omega_2$,
where the GRB photon has energy $\omega_1$, and the average energy of
CMB photons can be obtained from blackbody radiation by dividing the
energy density with photon number density:
$\omega_2=\frac{u(T)}{n(T)}=\frac{\frac{48\pi\zeta(4)}{c^3h^3}(k_BT)^4}
{\frac{16\pi\zeta(3)}{c^3h^3}(k_BT)^3}=\frac{3\zeta(4)}{\zeta(3)}k_BT
=2.33\times10^{-4}\;\frac{\mathrm{eV}}{\mathrm{K}}\times T$.
Hence we obtain a typical energy of the process as
$\sqrt{\langle s\rangle}=\sqrt{\omega_{1}/\mathrm{GeV}}\times
1.1\;\mathrm{keV}$.

\newcommand{\atitle}[1]{\emph{#1},}
\newcommand{\jref}[2]{\href{https://doi.org/#2}{#1}}
\newcommand{\arXiv}[2]{\href{https://arxiv.org/abs/#1}
{\texttt{arXiv:#1 [#2]}}}
\newcommand{\arXivOld}[1]{\href{https://arxiv.org/abs/#1}{\texttt{#1}}}

\end{document}